\documentclass[letter]{aa} % for the letters 

\usepackage{graphicx}
\usepackage{txfonts}
\usepackage{natbib}

\begin{document}

\title{Chemical similarities between Galactic bulge and local thick disk red giant stars}

\titlerunning{Chemical similarities between the Galactic bulge and thick disk}
\authorrunning{Mel\'{e}ndez et al.}

\author{
J. Mel\'{e}ndez\inst{1,2} \and
M. Asplund\inst{3} \and
A. Alves-Brito\inst{4} \and
K. Cunha \inst{5,6} \and
B. Barbuy \inst{4} \and
M.S. Bessell \inst{2} \and
C. Chiappini \inst{7,8} \and
K.C. Freeman \inst{2} \and
I. Ram\'{\i}rez \inst{9} \and
V.V. Smith \inst{5} \and
D. Yong\inst{2} 
          }

%\offprints{J. Mel\'{e}ndez (\email{jorge@mso.anu.edu.au}}

\institute{
Centro de Astrof\'{\i}sica da Universidade do Porto, Rua das Estrelas, 4150-762 Porto, Portugal \and
Research School of Astronomy and Astrophysics,
The Australian National University, Cotter Road, Weston, ACT 2611, Australia \and
Max Planck Institute for Astrophysics,
Postfach 1317, 85741 Garching, Germany \and
Universidade de S\~{a}o Paulo, IAG, Rua do Mat\~{a}o 1226, 
Cidade Universit\'{a}ria, S\~{a}o Paulo 05508-900, Brazil \and
National Optical Astronomy Observatory, Casilla 603, La Serena, Chile \and
On leave from Observat\'orio Nacional, Rio de Janeiro, Brazil \and
Geneva Observatory, Ch. des Maillettes 51, 1290 Sauverny, Switzerland \and
OAT/INAF, Via Tiepolo 11, Trieste 34131, Italy \and
McDonald Observatory and Department of Astronomy, University of Texas, 
RLM 15.306, Austin, TX 78712-1083, USA
             }

\date{Received: January 15, 2008; accepted: April 18, 2008}

\abstract
% context heading (optional)
{The evolution of the Milky Way bulge and its relationship
with the other Galactic populations is still poorly understood.
The bulge has been suggested to be either a merger-driven classical bulge or
the product of a dynamical instability of the inner disk.}
% aims heading (mandatory)
{To probe the star formation history, the initial mass
function and stellar nucleosynthesis of the bulge, we performed
an elemental abundance analysis of bulge red giant stars. 
We also completed an identical study of local thin disk, thick disk and halo giants 
to establish the chemical differences and 
similarities between the various populations.}
% methods heading (mandatory)
{High-resolution infrared spectra of 19 bulge giants and 49 comparison
giants in the solar neighborhood were acquired with Gemini/Phoenix.
All stars have similar stellar parameters but cover a broad range in metallicity. 
A standard 1D  local thermodynamic equilibrium analysis yielded the 
abundances of C, N, O and Fe.
A homogeneous and differential analysis of the bulge, halo, thin disk and thick disk stars 
ensured that systematic errors were minimized.}
% results heading (mandatory)
{We confirm the well-established differences for [O/Fe] (at a given
metallicity) between the local thin and thick disks. 
For the elements investigated, we find no chemical distinction between
the bulge and the local thick disk, 
which is in contrast
to previous studies relying on literature values for disk dwarf stars 
in the solar neighborhood.}
% conclusions heading (optional) 
{Our findings  suggest that the bulge and local thick disk 
experienced similar, but not necessarily shared, chemical evolution histories. 
We argue that their formation timescales, star formation rates and
initial mass functions were similar. 
%It is important to extend our abundance analysis to the inner regions of the 
%Galactic disks in order to establish a causal relationship between the
%two stellar populations.
}


\maketitle
%
%________________________________________________________________

\section{Introduction}

Despite its prominent role
in the formation and evolution of the Galaxy, the bulge is the least well-understood 
stellar population in the Milky Way.
Two main formation scenarios have been proposed to explain
bulges of spiral galaxies 
\citep[see][for a review]{2004ARA&A..42..603K}. %(Kormendy \& Kennicutt 2004). 
In classical bulges, most
stars are formed during an early phase of intensive star
formation following collapse of the proto-galaxy
and subsequent mergers, as predicted in a cold dark matter cosmology.
Boxy or peanut-shaped bulges develop through dynamical 
instability of an already established inner disk.
%, as demonstrated by N-body simulations.
The nature of the Galactic bulge is still unknown since
its generally old and metal-rich stellar population
\citep[e.g.][]{2006A&A...457L...1Z, 2006ApJ...636..821F}
%(e.g. Zoccali et al. 2003; Fulbright et al. 2006)
is consistent with a classical bulge while its boxy shape
is indicative of formation by dynamical instability.

Compared to the other Galactic populations, our
understanding of the chemical properties
of the bulge and its evolution with time 
is still sketchy due to the large distance and
high visual extinction. 
Following pioneering work on 4\,m-class telescopes
\citep[e.g.][]{1994ApJS...91..749M},
%(e.g. McWilliam \& Rich 1994), 
the situation has, however, improved based on observations with 
8-10\,m telescopes
\citep[e.g.][]{2003A&A...411..417M, 2005ApJ...634.1293R,
2006A&A...457L...1Z, 2006ApJ...636..821F, 
2007ApJ...661.1152F, 2006ApJ...651..491C,
2007A&A...465..799L}.
%(Mel\'{e}ndez et al. 2003; Rich \& Origlia 2005;
%Fulbright et al. 2006, 2007; 
%Cunha \& Smith 2006; Zoccali et al. 2006; Lecureur et al. 2007).
High-resolution multi-object and 
infrared (IR) spectrographs have also aided the cause, such
that the number of red giants in the bulge 
with detailed abundance information based on
high-resolution spectroscopy now exceeds 100.

\begin{figure*}
\centering
\includegraphics{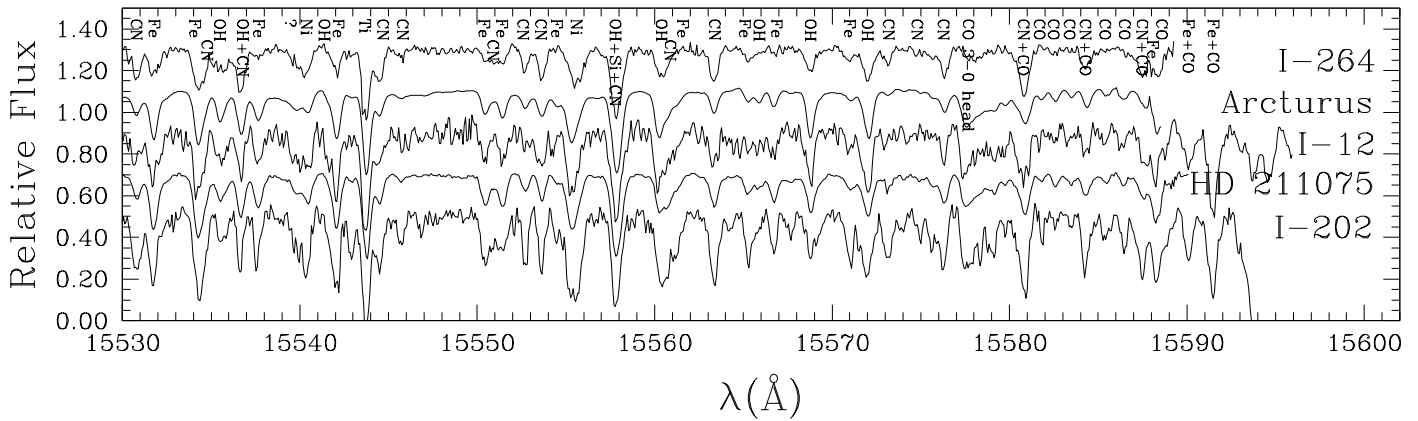}
\caption{Observed Phoenix spectra of selected bulge giants as well as 
thick (Arcturus = HD124897) and thin (HD211075) disk stars.}
\label{f:spectrum}
 \end{figure*}

Of particular interest are the abundances of $\alpha$-elements (such as O and Mg)
as a function of metallicity,
since they provide crucial information about the star formation
history and initial mass function 
\citep{1980FCPh....5..287T, 2007A&A...467..123B}.
%(Tinsley 1980; Ballero et al. 2007).
In regions of rapid star formation rates -- as expected
for the Galactic bulge -- higher metallicities are reached
before the contributions from thermonuclear supernovae (SNe\,Ia) 
are reflected in the abundance ratios.
The $\alpha$-element abundances should then remain
elevated compared to those parts of the Galaxy 
experiencing a more modest star formation rate.
A tell-tale signature is therefore $[\alpha / {\rm Fe}] > 0.0$
at high metallicities ($[{\rm Fe/H}] \approx 0$)
as a result of nucleosynthesis in core collapse supernovae (SNe\,II). 
A conclusion of most bulge studies is that indeed
the $\alpha$-elements seem to be overabundant relative to both the
thin and the thick disks in the solar neighborhood,
implying a short formation timescale of the bulge
\citep{2007ApJ...661.1152F, 2007A&A...465..799L}.
%(Lecureur et al. 2007; Fulbright et al. 2007).

In this {\em Letter} we revisit the issue
of the chemical differences between the bulge and the thin/thick
disks based on Gemini/Phoenix high-resolution IR spectra 
of 19 K giants in Baade's window as well as a sample
of nearby thin and thick disk giants.
The unique aspect of our study is that we conduct
a homogeneous differential analysis of stars with similar 
stellar parameters to minimize systematic errors.
The novel result of our careful analysis is that we find no significant
abundance differences for the elements studied (C, N and O) between
the bulge and the local thick-disk stars.

\section{Observations and abundance analysis}
\label{s:analysis}

High-resolution ($R \equiv \lambda/\Delta\lambda = 50,000$) 
IR spectra of 19 K giants in the Baade's window of the bulge 
were taken with the Phoenix spectrograph 
\citep{2003SPIE.4834..353H} %(Hinkle et al. 2003) 
on the 8\,m Gemini-South telescope. 
%All of our bulge giants have previously been analysed using optical spectra by 
%Fulbright et al. (2006, 2007).
The Phoenix spectra of five of these bulge giants were previously
analyzed by \citet{2006ApJ...651..491C}. %Cunha \& Smith (2006)
Importantly, comparison K giants of the thin disk (24 stars), thick disk 
(21 stars) and halo (4 stars) in the solar neighborhood were observed using 
the same instrument on Gemini-South as well as
the 2.1m and 4m Kitt Peak telescopes.
The stars were selected to cover the metallicity
range $-1.5 \le [{\rm Fe/H}] \le +0.5$. 
The assignment of population membership to the comparison sample
was based on UVW velocities 
\citep{2004A&A...415..155B, 2006MNRAS.367.1329R}. 
%(Bensby et al. 2004; Reddy et al. 2006) 
All IR spectra were obtained using the same instrumental setup centered
on 1.5555\,$\mu$m (Fig. \ref{f:spectrum}).

We also acquired high-resolution ($R = 60,000$) optical spectra 
to check the derived stellar parameters employing Fe\,{\sc i} and 
Fe\,{\sc ii} lines. 
The disk and halo stars were observed using the MIKE spectrograph 
on the Clay 6.5\,m Magellan telescope and the 
2dcoud\'e  spectrograph %(Tull et~al. 1995) 
on the 2.7\,m Harlan J. Smith 
telescope at McDonald Observatory. For the bulge
stars, we rely on the equivalent widths measured by 
\citet{2006ApJ...636..821F,  2007ApJ...661.1152F}
%Fulbright et al. (2006, 2007) 
using the HIRES spectrograph on the Keck-I 10\,m telescope.
Our observations were acquired between September 1999 and
October 2007. 

Both the optical and infrared spectra were reduced homogeneously
with IRAF; we refer the reader to 
\citet{2003A&A...411..417M}, %Mel\'endez et al. (2003), 
\citet{2006ApJ...639..918Y} %Yong et al. (2006)
and \citet{2007ApJ...669L..89M} %Mel\'endez \& Ram\'{\i}rez (2007)
for details on the data reduction of 
Phoenix, MIKE, and 2dcoude spectra, respectively. 
%A detailed description will be given in 
%Mel\'endez et al. (2008, in preparation). 
The signal-to-noise ratio (S/N) of the reduced spectra ranges from $S/N \approx 100$ per spectral 
resolution element, for the IR spectra of the bulge giants, to 
several hundreds per pixel, for both the optical and IR 
spectra of the bright disk and halo K giants.

Photometric temperatures were obtained using optical and infrared
colors and the infrared flux method $T_{\rm eff}$-scale of 
\citet{2005ApJ...626..465R}. %Ram\'{\i}rez \& Mel\'endez (2005) 
Reddening for the
bulge stars was estimated from extinction maps 
\citep{1996ApJ...460L..37S} %(Stanek 1996) 
while for the comparison samples both extinction maps
and Na\,{\sc i} D ISM absorption lines were adopted  \citep{2006ApJ...642.1082M}.
The stellar surface gravities 
were derived from improved Hipparcos parallaxes 
\citep{2007A&A...474..653V} %(van Leeuwen 2007) 
using bolometric corrections from 
\citet{1999A&AS..140..261A} %Alonso et al. 1999
for the relatively nearby disk and halo stars and assuming a distance
of 8\,kpc for the bulge giants. 
In addition, Yonsei-Yale 
\citep{2004ApJS..155..667D} %(Demarque et al. 2004) 
and Padova isochrones 
\citep{2006A&A...458..609D} %(da Silva et al. 2006) 
were employed to
determine evolutionary gravities.
%; the $\log g$ values obtained from 
%isochrones required small zero-point corrections of -0.06 dex and +0.04\,dex,
%respectively, to be on the same scale as the Hipparcos-based results. 
Tests of the ionization and excitation balances of Fe\,{\sc i} and 
Fe\,{\sc ii} lines revealed that the photometric 
stellar parameters of only six stars required significant adjustments;
given the remaining uncertainties in both the photometric and spectroscopic
parameters the overall agreement for the remaining stars is encouraging.
The adopted stellar parameters are given in Table \ref{t:stars}.
We estimate that our stellar parameters have typical uncertainties
of $\Delta T_{\rm eff} \approx \pm 75$\,K, $\Delta \log g \approx \pm 0.3$
and $\Delta {\rm v_t} \approx 0.2$ km s$^{-1}$, which 
translates into abundance errors for Fe, C, N and O of 0.03, 0.11, 0.11 and
0.14\,dex, respectively, when the errors are added in quadrature; 
the errors in [X/Fe] are almost identical.
%\citep{2003A&A...411..417M}. %Mel\'endez et al. (2003)
As discussed below, these uncertainties are probably too conservative;
an uncertainty in [O/Fe] of 0.10\,dex was adopted.

The Phoenix spectral window contains several useful lines 
for abundance purposes.
The stellar C, N, O and Fe abundances were obtained from
spectrum synthesis of a number of  CO, CN, OH and Fe\,{\sc i} lines
using MOOG 
\citep{1973PhDT.......180S}; %(Sneden 1973) 
it is important to be able
to derive the C, N and O abundances consistently for each
star given the interdependencies in the respective molecular balance
and corresponding line strengths.
%In addition Ti and Ni abundances have been determined although
%we place lower weight on in particular the Ti results given 
%that they are based on a singular, less than ideal Ti\,{\sc i} line. 
The same transition probabilities for the atoms and molecules were
applied to both the bulge and comparison samples
\citep{2003A&A...411..417M}. %Mel\'endez et al. (2003).
In the present work, we employed both
Kurucz models with convective overshooting 
%\citet{1993ASPC...44...87K} %
\citep{1997A&A...318..841C} %Castelli et al. (1997, A\&A, 318, 841)  
and specially calculated MARCS 
\citep{2003ASPC..288..331G} %(Gustafsson et al. 2003) 
1D hydrostatic model atmospheres.
For the MARCS models, both $\alpha$-enhanced ($[\alpha/{\rm Fe}]=+0.2$
and $+0.4$) and scaled-solar abundances models were constructed;
for the Kurucz' models, adjustments of [Fe/H] 
were applied to simulate the effects of $\alpha$-enhancement
on the model atmospheres: $\Delta  [{\rm Fe/H}] = \log (0.64 \times  10^{[\alpha/{\rm Fe}]} + 0.36)$
\citep{1993ApJ...414..580S}. % (Salaris et al. 1993, ApJ, 414, 580).
Our final results 
%presented in Table \ref{t:abund} 
are based on the MARCS models with appropriate compositions. 
The effects of failing to account for the variations in $[\alpha/{\rm Fe}]$
can be substantial: a difference of $[\alpha/{\rm Fe}]$ = +0.2 in the model
atmosphere corresponds to a change of $\approx$ +0.1 dex in the derived [O/Fe].
Otherwise, the agreement between MARCS and Kurucz models
is very good: the mean differences for [Fe/H], [C/Fe], [N/Fe] and [O/Fe]
between MARCS and Kurucz models are only +0.01, +0.01, +0.02 and +0.04 dex, respectively. 
The corresponding solar abundances for the MARCS suite of models are  
$\log \epsilon_{\rm C} = 8.42$,
$\log \epsilon_{\rm N} = 7.82$, 
$\log \epsilon_{\rm O} = 8.72$ and
$\log \epsilon_{\rm Fe} = 7.48$, similar to the 3D-based values provided by
\citet{2005ASPC..336...25A}. %Asplund et al. (2005).
For the bulge stars in common, the differences in [O/Fe] between us and  
 \citet{2007ApJ...661.1152F} %Fulbright et al. (2007) 
is $+0.03 \pm 0.13$\,dex.

While no predictions of 3D hydrodynamical models are available
for precisely our parameter space of interest
\citep{2005ARA&A..43..481A}, %(Asplund 2005)
we note that, according to simulations of slightly less evolved 
red giants, the 3D abundance corrections for the species
considered herein are expected to be modest: 
$|\Delta \log \epsilon| \la 0.1$\,dex at $[{\rm Fe/H}] \ge 0$,
although slightly larger at the lowest metallicities of our targets
\citep{2007A&A...469..687C}. %(Collet et al. 2007).
Given the similarity in parameters between the
bulge and disk giants, the {\em relative} abundance ratio
differences will be significantly smaller.
% and thus inconsequential for our conclusions.

\section{Results}

\begin{figure}
\centering
\includegraphics[width=\hsize]{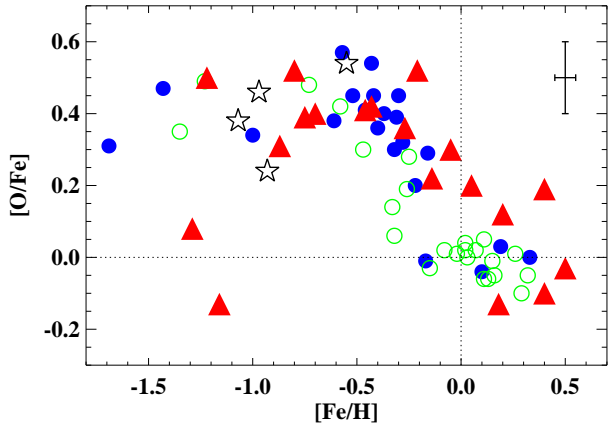}
\caption{The derived [O/Fe] ratios, as a 
function of [Fe/H], for the bulge (red triangles), thick disk (blue solid circles),
thin disk (open green circles) and halo giants (stars). A typical error bar is shown.
Note the similarities between the bulge and thick disk trends for $[{\rm Fe/H}] < -0.2$.
}
\label{f:ofe}
 \end{figure}

\begin{figure}
\centering
\includegraphics[width=\hsize]{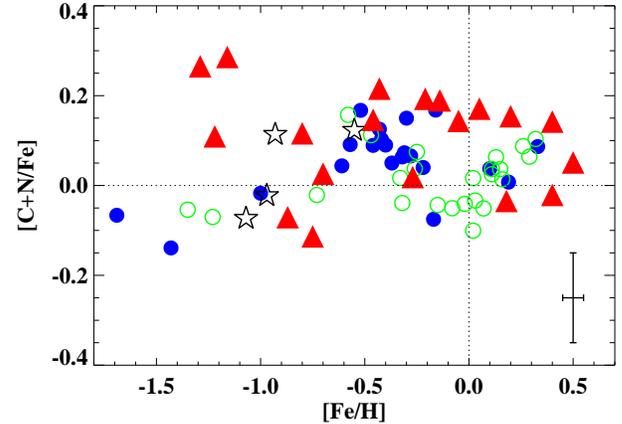}
\caption{The derived [C+N/Fe] ratios as a 
function of [Fe/H] for the bulge (red triangles), thick disk (blue solid circles),
thin disk (open green circles) and halo giants (stars). A typical error bar is shown.
%The fact that all stars
%have [C+N/Fe]\,$\approx 0.0$ suggests that while CN-cycling may have
%occurred, significant amounts of CNO-cycled gas has not been
%dredged up to the stellar atmospheres.
}
\label{f:cnfe}
 \end{figure}

Our results in terms of [O/Fe] are shown in Fig. \ref{f:ofe}.
% from which several noteworthy conclusions can be drawn. 
One concern is whether
CNO-cycled material has been dredged-up to the surface in our giants. 
It is clear that many of the stars' atmospheres contain CN-cycled gas.
As demonstrated in Fig. \ref{f:cnfe}, however,
the [C+N/Fe] ratio remains roughly constant at solar values 
as a function of  metallicity for both the bulge and disk giants,
which implies that intrinsic 
CNO-cycled rest-products have not been brought up to the surface. 
We therefore believe that the measured O abundances are indeed a proper
reflection of the pristine contents the stars were born with.
We note that 
\citet{2007ApJ...661.1152F}
%Fulbright et al. (2007) 
argued that the two O-deficient stars (I-264 and IV-203) 
at $[{\rm Fe/H}] \approx -1.2$ have experienced 
envelope H-burning reminiscent of that believed to be responsible 
for the O-Na correlations in globular clusters based on the stars' 
high Na and Al abundances. 
It is then surprising that their [C+N/Fe] ratios are normal or only marginally higher as O-depletion
in globular cluster stars is associated with increased [N/Fe] ratios due to CNO-cycling 
\citep[][and references therein]{2004ARA&A..42..385G}. %(Gratton et al. 2004).
In any case, we believe that the O abundances of these two stars
do not reflect the typical bulge composition. 

In agreement with previous findings 
\citep{2006A&A...457L...1Z, 2007ApJ...661.1152F, 2007A&A...465..799L},
%(Zoccali et al. 2006; Lecureur et al. 2007; Fulbright et al. 2007), 
the bulge [O/Fe] trend goes from typical halo values to roughly solar or below at
$[{\rm Fe/H}]>0$. The break in [O/Fe] at $[{\rm Fe/H}] \approx -0.3$
implies that both SNe\,II and SNe\,Ia have contributed and therefore
that the formation of the bulge proceeded over several 100\,Myr.
The plateau in [O/Fe] for the bulge extends to higher metallicities than for
the thin disk, implying a higher star formation rate. 
Furthermore, we confirm the by now well-established chemical
distinctions of the Galactic thin and thick disks locally, in the sense
that for a given $[{\rm Fe/H}]$ the latter is more over-abundant
in $\alpha$-elements 
\citep{2004A&A...415..155B, 2006MNRAS.367.1329R}.
%(Bensby et al. 2004; Reddy et al. 2006). 
We also seem to detect a similar knee in [O/Fe]  for the local thick disk  as advocated by 
\citet{2004A&A...415..155B}, %Bensby et al. (2004), 
although this is based on only a handful of stars.
The existence of this SNe\,Ia signature has been questioned
by other studies of thick-disk stars 
\citep{2006MNRAS.367.1329R, 2007A&A...465..271R}.
%(Reddy et al. 2006; Ram\'{\i}rez et al. 2007). 
The interpretation largely hinges on whether or not the 
kinematically selected local thick-disk stars at  $[{\rm Fe/H}] \ga -0.2$
truly belong to this population or are just part of the high-velocity tail 
of the thin disk. 
As also seen from the four such thick-disk stars included in our sample,
there is no chemical differentiation at these metallicities
between the thin and thick disk, if indeed
the thick disk extends to these high [Fe/H]. 
Our limited number of thick-disk stars is insufficient to draw
firm conclusions to these questions. 
Likewise, the four thin-disk stars with $[{\rm Fe/H}] < -0.5$ 
may belong to the thick disk since our classification 
assumes a metallicity-independent thick/thin disk stellar fraction of 10\%
\citep{2004A&A...415..155B}. %Bensby et al. (2004).

The most surprising inference of our study comes from a comparison of
our bulge and local thick-disk samples. 
In contrast to previous works on the topic
\citep{2006A&A...457L...1Z, 2007ApJ...661.1152F, 2007A&A...465..799L},
%(Zoccali et al. 2006; Lecureur et al. 2007; Fulbright et al. 2007), 
we find that the two
populations are indistinguishable in their abundance patterns for the elements considered
(C, N and O) up to $[{\rm Fe/H}]=-0.2$,
i.e. to the metallicity range where the thick disk is unambiguously identified. 
A linear fit to the bulge data is $[{\rm O/Fe}] = 0.41 - 0.02 \times [{\rm Fe/H}]$ while for
the thick disk it is $[{\rm O/Fe}] = 0.39 - 0.01 \times [{\rm Fe/H}]$, both with 
a scatter of $\sigma = 0.09$\,dex; the mean difference between the bulge
fit and the thick disk data is $0.03 \pm 0.09$\,dex. The comparison implies that
the real abundance errors probably do not exceed $0.10$\,dex rather than the higher estimates 
given in Sect. \ref{s:analysis}.
The metallicity of the bulge extends to significantly higher [Fe/H] than that, which, as explained above,
remains to be convincingly demonstrated for the thick disk. 
Rather than having had a significantly higher star formation rate as normally argued,
the conclusion is thus that the bulge did not differ noticeably from the local thick disk 
in this respect. Furthermore, our observations suggest that the 
initial mass function for the two populations were similar.
The nearly identical  [O/Fe] trends do not necessarily imply a causal
relationship between the bulge and local thick disk. 
Such a relationship has been proposed for other spiral galaxies
\citep[e.g.][]{1981A&A....95..105V} % (van der Kruit \& Searle 1981)
and remains an intriguing possibility based on our observations. 

In the classical bulge scenario one does not expect any direct relationship
between the bulge and either of the disk populations. Since the thin disk
had a long formation timescale and experienced a smaller star formation rate, 
it is unsurprising that [O/Fe] differs between the solar neighborhood thin disk
and the bulge. 
If both the bulge and the thick disk formed on short timescales as well as had
similar initial mass function and star formation histories, 
the abundance pattern would be similar even
if the two populations lack a physical connection.
In the disk instability formation model for the Galactic bulge, the bulge stellar population 
consists largely of stars from the inner disk that have been heated dynamically
through the instabilities associated with bar formation and buckling.
According to N-body simulations, the bar and consequently the bulge appears
only after a few Gyr, when the thick disk was already in place according
to the observed stellar ages of $\ga 9$\,Gyr
\citep{2007ApJ...663L..13B}. %Bensby et al. (2007). 
However, it is important to distinguish between the ages of the stars and the ages
of the structures that they inhabit. The stars of the bulge and thick disk 
appear to be very old. If the bulge and thick disk both formed dynamically
(e.g. via instabilities or minor merger heating of the early thin disk),
then the bulge and thick-disk stars may well
be significantly older than the bulge and thick-disk structures, and the bulge-thick disk
similarities which we have uncovered would then not be surprising.
The differences between the bulge and our
thin-disk stars do not support a causal link between the two,
but may simply reflect the Galactic radii probed. 
It is not known empirically how [O/Fe] varies in the inner thin disk but
models constructed to reproduce the Galactic abundance gradient
imply that [O/Fe] at a Galactocentric radius of 4\,kpc is
only $\approx 0.02$\,dex higher than at the solar location of 8\,kpc
\citep{2000ApJ...528..711C, 2007A&A...462..943C}.
%(Chiappini et al. 2001, Cescutti et al. 2007)
This suggests that there should be a significant difference
in [O/Fe] between the bulge and the thin disk in the transition
region in the inner parts of the Galaxy. 
It remains to be seen whether this conclusion holds also
observationally.

We conclude the discussion by briefly arguing why our 
detected similarities and differences between the populations are robust findings. 
Since we are primarily interested in
studying the relative differences between our bulge, disk and halo samples,
the zero-points adopted for our stellar parameters does not
matter much, since all stars, regardless of population, have 
comparable properties determined similarly.
Furthermore, both bulge and comparison samples were analysed
using the same set of lines.
We believe that not adhering to these principles is the main reason
previous studies 
\citep[e.g.][]{2006A&A...457L...1Z, 2007ApJ...661.1152F, 2007A&A...465..799L}
%(e.g. Fulbright et al. 2006, 2007; Zoccali et al. 2006; Lecureur et al. 2007)
reached different conclusions, by misleadingly comparing their bulge giant
results with literature values for main sequence and turn-off disk stars 
in the solar neighborhood (Bensby et al. 2005; Reddy et al. 2006). 
Even the particular choice of solar abundances for the normalization of
the stellar results in different works can introduce errors
as large as the purported abundance differences between the various
stellar populations. 
We circumvent these problems here by adopting a differential
analysis of the bulge and disk giants.

\section{Concluding remarks}

Three obvious follow-up investigations to the present study are urgently needed.
The first is to obtain larger samples of bulge and thick-disk giants, 
especially in the critical metallicity range $-0.5 \le [{\rm Fe/H}] \le +0.0$, to 
confirm that our conclusions regarding the chemical similarities remain unaltered. 
Secondly, the analysis should be extended to a determination of additional
elements; here Mg is particularly desirable, since
\citet{2007ApJ...661.1152F} %Fulbright et al. (2007)
and \citet{2007A&A...465..799L} %Lecureur et al. (2007)
argue that it shows pronounced differences of about 0.2\,dex 
between the bulge giants and local thick-disk dwarfs. 
Finally, one needs to have a comparison sample from the inner disk
analysed in a homogeneous way. 
Do the chemical similarities between the bulge and the thick disk remain
when moving from the solar neighborhood to the inner regions of the Galaxy?
Is the Galactic abundance gradient sufficient to make
the inner thin disk indistinguishable from the bulge? 
If so, it would argue in favor of a disk instability rather than a merger origin for the bulge. 
Until the launch of GAIA, it will be difficult to differentiate kinematically 
between the thin and thick disks at such large distances,
but the chemical signatures may be sufficient since 
\citet{2007A&A...465..271R} %Ram\'{\i}rez et al. (2007) 
have shown that there is no intermediate disk population
at $[{\rm Fe/H}] \approx -0.3$, locally.
More work is clearly needed to establish the causal relationships, if any, 
between the bulge and the thin and thick disks in the inner regions of the Galaxy.

\begin{acknowledgements}
We thank S. Schuler and K. Hinkle for assisting with 
carrying out some of the Gemini/Phoenix observations.
Based on observations obtained at the Gemini Observatory, which is operated by the
AURA, Inc., %Association of Universities for Research in Astronomy, Inc., 
under a cooperative agreement
with the NSF on behalf of the Gemini partnership: 
the NSF %National Science Foundation 
(United States), 
the STFC %Science and Technology Facilities Council 
(United Kingdom), 
the NRC %National Research Council 
(Canada), CONICYT (Chile), 
the ARC %Australian Research Council
(Australia), CNPq (Brazil) and SECYT (Argentina).
This paper uses data obtained with the Phoenix infrared spectrograph, developed
and operated by 
NOAO. %National Optical Astronomy Observatory.
This work has been supported by 
ARC %Australian Research Council
(DP0588836), 
ANSTO (06/07-0-11) and
NSF %National Science Foundation 
(AST 06-46790).
\end{acknowledgements}

\bibliography{bulge}

\newpage

\begin{table}
\caption{Stellar parameters and derived abundances}
\label{t:stars}
\centering          
\begin{tabular}{lcccccc}      
\hline\hline                
Star & $T_{\rm eff}$ & $\log g$ & [Fe/H] & [C/Fe] & [N/Fe] & [O/Fe] \\
\hline    
{\bf Bulge:} &&&&&&\\                    
        I012 &  4237 &  1.61 & -0.43 &  0.05 &  0.57 &  0.42\\
        I025 &  4370 &  2.28 &  0.50 & -0.27 &  0.54 & -0.03\\
        I141 &  4356 &  1.93 & -0.21 &  0.07 &  0.49 &  0.52\\
        I151 &  4434 &  1.73 & -0.80 &  0.05 &  0.31 &  0.52\\
        I156 &  4296 &  1.60 & -0.70 & -0.04 &  0.22 &  0.40\\
        I158 &  4426 &  2.68 & -0.14 &  0.12 &  0.39 &  0.22\\
        I194 &  4183 &  1.67 & -0.27 & -0.19 &  0.42 &  0.36\\
        I202 &  4252 &  2.07 &  0.20 & -0.03 &  0.53 &  0.12\\
        I264 &  4046 &  0.68 & -1.16 & -0.33 &  0.89 & -0.13\\
        I322 &  4255 &  1.90 & -0.05 &  0.04 &  0.41 &  0.30\\
       II033 &  4230 &  1.37 & -0.75 & -0.37 &  0.33 &  0.39\\
      III152 &  4143 &  1.51 & -0.46 &  0.08 &  0.34 &  0.41\\
      IV003 &  4500 &  1.85 & -1.22 & -0.17 &  0.57 &  0.50\\
      IV072 &  4276 &  2.13 &  0.18 & -0.20 &  0.32 & -0.13\\
      IV167 &  4374 &  2.44 &  0.40 & -0.24 &  0.39 & -0.10\\
     IV203 &  3815 &  0.35 & -1.29 & -0.58 &  0.91 &  0.08\\
     IV325 &  4353 &  2.35 &  0.40 & -0.01 &  0.48 &  0.19\\
      IV329 &  4153 &  1.15 & -0.87 & -0.25 &  0.30 &  0.31\\
      BW96 &  4050 &  1.20 &  0.05 & -0.08 &  0.61 &  0.20\\
{\bf Halo:} &&&&&&\\
    HD041667 &  4581 &  1.80 & -1.07 & -0.20 &  0.23 &  0.38\\
    HD078050 &  4951 &  2.54 & -0.93 &  0.14 & -0.01 &  0.24\\
    HD114095 &  4794 &  2.68 & -0.55 &  0.11 &  0.17 &  0.54\\
    HD206642 &  4372 &  1.57 & -0.97 & -0.13 &  0.25 &  0.46\\
{\bf Thick disk:} &&&&&&\\
    HD008724 &  4577 &  1.49 & -1.69 & -0.40 &  0.43 &  0.31\\
    HD077236 &  4427 &  2.01 & -0.57 &  0.04 &  0.25 &  0.57\\
    HD030608 &  4620 &  2.39 & -0.22 & -0.13 &  0.40 &  0.20\\
    HD107328 &  4417 &  2.01 & -0.30 &  0.11 &  0.28 &  0.45\\
    HD023940 &  4762 &  2.61 & -0.32 & -0.08 &  0.39 &  0.30\\
    HD032440 &  3941 &  1.15 & -0.17 & -0.25 &  0.29 & -0.01\\
    HD037763 &  4630 &  3.15 &  0.33 & -0.07 &  0.43 &  0.00\\
    HD040409 &  4746 &  3.20 &  0.10 & -0.07 &  0.31 & -0.04\\
    HD077729 &  4127 &  1.48 & -0.43 &  0.08 &  0.27 &  0.54\\
    HD080811 &  4900 &  3.28 & -0.37 & -0.05 &  0.31 &  0.40\\
    HD083212 &  4480 &  1.55 & -1.43 & -0.43 &  0.33 &  0.47\\
    HD099978 &  4678 &  2.07 & -1.00 & -0.03 &  0.03 &  0.34\\
    HD107773 &  4891 &  3.28 & -0.31 & -0.02 &  0.32 &  0.39\\
    HD119971 &  4093 &  1.36 & -0.61 & -0.06 &  0.31 &  0.38\\
    HD124897 &  4280 &  1.69 & -0.52 &  0.15 &  0.23 &  0.45\\
    HD130952 &  4742 &  2.53 & -0.28 & -0.01 &  0.28 &  0.32\\
    HD136014 &  4774 &  2.52 & -0.40 &  0.03 &  0.27 &  0.36\\
    HD145148 &  4851 &  3.67 &  0.19 & -0.10 &  0.28 &  0.03\\
    HD180928 &  4092 &  1.48 & -0.42 &  0.06 &  0.24 &  0.45\\
    HD203344 &  4666 &  2.53 & -0.16 &  0.02 &  0.50 &  0.29\\
    HD219615 &  4833 &  2.51 & -0.46 &  0.02 &  0.29 &  0.41\\
{\bf Thin disk:} &&&&&&\\
    HD000787 &  4020 &  1.36 &  0.07 & -0.24 &  0.33 &  0.02\\
    HD005268 &  4873 &  2.54 & -0.47 & -0.04 &  0.45 &  0.30\\
    HD005457 &  4631 &  2.67 &  0.03 & -0.19 &  0.31 &  0.00\\
    HD018884 &  3731 &  0.73 &  0.02 & -0.33 &  0.32 &  0.02\\
    HD029139 &  3891 &  1.20 & -0.15 & -0.20 &  0.30 & -0.03\\
    HD029503 &  4616 &  2.67 &  0.11 & -0.22 &  0.48 &  0.05\\
    HD050778 &  4034 &  1.40 & -0.26 & -0.07 &  0.31 &  0.19\\
    HD099648 &  4837 &  2.24 & -0.02 & -0.23 &  0.34 &  0.01\\
    HD100920 &  4788 &  2.58 & -0.08 & -0.23 &  0.32 &  0.02\\
    HD115478 &  4272 &  2.00 &  0.02 & -0.08 &  0.27 &  0.04\\
    HD116976 &  4691 &  2.49 &  0.11 & -0.09 &  0.31 & -0.06\\
    HD117818 &  4802 &  2.59 & -0.25 & -0.16 &  0.50 &  0.28\\
    HD128188 &  4657 &  2.03 & -1.35 & -0.12 &  0.14 &  0.35\\
    HD132345 &  4400 &  2.34 &  0.29 & -0.12 &  0.44 & -0.10\\
    HD142948 &  4820 &  2.24 & -0.73 & -0.24 &  0.39 &  0.48\\
    HD171496 &  4975 &  2.40 & -0.58 &  0.14 &  0.22 &  0.42\\
    HD172223 &  4471 &  2.46 &  0.26 & -0.02 &  0.36 &  0.01\\
    HD175219 &  4720 &  2.44 & -0.32 & -0.20 &  0.31 &  0.06\\
    HD186378 &  4566 &  2.42 &  0.16 & -0.09 &  0.28 & -0.05\\
    HD187195 &  4405 &  2.36 &  0.13 & -0.06 &  0.36 & -0.06\\
    HD210295 &  4738 &  1.88 & -1.23 & -0.11 &  0.06 &  0.49\\
    HD211075 &  4305 &  1.76 & -0.33 & -0.01 &  0.11 &  0.14\\
    HD214376 &  4586 &  2.55 &  0.15 & -0.06 &  0.29 & -0.01\\
    HD221148 &  4663 &  3.22 &  0.32 & -0.04 &  0.43 & -0.05\\
\hline                                 
\end{tabular}
\end{table}

\end{document}